\documentclass{ws-procs975x65}

\begin{document}

\title{Production Rate of Second KK Gauge Bosons \\ 
        in UED Models at LHC}

\author{Shigeki Matsumoto}

\address{Department of Physics, University of Toyama, \\ Toyama 930-8555, Japan \\
E-mail: smatsu@sci.u-toyama.ac.jp}

\author{Joe Sato$^*$ and Masato Yamanaka$^\dagger $}

\address{Department of Physics, Saitama University, \\
     Shimo-okubo, Sakura-ku, Saitama, 338-8570, Japan\\
$^*$E-mail: joe@phy.saitama-u.ac.jp \\
$^\dagger$E-mail: masa@krishna.phy.saitama-u.ac.jp}

\author{Masato Senami}

\address{Department of Micro Engineering,
      Kyoto University, \\ Kyoto 606-8501, Japan \\
E-mail: senami@me.kyoto-u.ac.jp}

\begin{abstract}
We calculate the production rates of the second
Kaluza-Klein (KK) photon $\gamma^{(2)}$ and  Z boson
$Z^{(2)}$ at the LHC including all significant processes in the minimal 
universal extra dimension (MUED) model. For discrimination of the
MUED model from other TeV scale models in
hadron collider experiments, $\gamma^{(2)}$ and $Z^{(2)}$ play a crucial
role. In order to discuss the discrimination and calculate their
production rates, we derive effective Lagrangian containing KK
number violating operators. We show that KK number violating processes
are extremely important for the compactification scale larger than 800
GeV. We find that, with an integrated luminosity of 100 fb$^{-1}$,
$\gamma^{(2)}$ and $Z^{(2)}$ are produced 10$^6$ - 10$^2$ for the 
compactification scale between 400 GeV and 2000 GeV.
\end{abstract}


\bodymatter

\section{Introduction}\label{sec1} 

Universal Extra Dimension (UED) models \cite{Appelquist:2000nn} are one 
of attractive candidates for new physics at TeV scale.
Among various UED models, the simplest and the most popular one is
called the Minimal UED (MUED) model. The MUED model is defined on the
five dimensional space-time, where the extra dimension is compactified
on an $S^1/Z_2$ orbifold. 
%
%
In the MUED model, only two parameters are
newly introduced to the SM. One is the compactification scale of the
extra dimension $1/R$, the inverse of the radius of $S^1$ circle, and
the other is the cutoff scale of the MUED model $\Lambda$. In this article, 
we take $\Lambda$ to be $\Lambda R = 20$.
In the MUED model, all SM fields can propagate into compactified extra
dimension, and fields with a fifth dimensional momentum $n/R$ behave 
as new heavy particles with a mass 
$m_n = \sqrt{ (n/R)^2 + m_{SM}^2 + \delta m_n^2 }~$ from the viewpoint 
of four dimensional field theory. Here analytical expression of the 
radiative corrections $\delta m_n$ are given in Ref. 2. 
These new particles are called Kaluza-Klein (KK) particles, $n$ is called 
the KK number ($n$ = 0 for SM particles, $n$ = 1, 2, ... for KK particles), 
and $m_{SM}$ represents the mass of the corresponding SM particle. They 
can give plausible explanations for the existence of dark matter 
\cite{Cheng:2002ej}, SM neutrino masses which are embedded in extended 
models \cite{Matsumoto:2006bf}, and so on.

Since the translational invariance along the extra dimension direction is broken 
due to the orbifolding, the fifth dimensional momentum (KK number) is no longer 
conserved. Nevertheless, the subgroup of the translational invariance remains 
unbroken, which is called the KK parity. Under the parity, particles with even 
(odd) KK number have plus (minus) sign, and the product of the sign is conserved 
in each process. Because of the KK parity, the lightest KK particle (LKP) is stable 
and provided as a candidate for dark matter. 
The relic abundance of the KK particle dark matter has been calculated 
\cite{Kakizaki:2005en}, and it turns out that $1/R$ should be in the range of 500 
GeV-1500 GeV. The MUED model, therefore, would be explored at the LHC.

The confirmation of the MUED model needs the discovery of 
KK particles. Though the LHC can produce KK particles, it is difficult to
confirm that they are indeed the KK particles, because new particles
predicted in various TeV scale models give quite similar signatures to
each other. Therefore it is very important to understand how they can be
identified as the KK particles. In this article, we discuss the discrimination 
of the MUED model from other TeV scale models.

An idea to discriminate the MUED model from other models has been proposed 
\cite{Datta:2005zs}. The essence of the discrimination is the discovery of the 
second KK particles. 
The discovery of the second KK particles suggests the existence of the MUED 
model, since their masses are peculiar which is almost equal to 
$2/R$ and $1/R$ is expected by the masses of the ``first KK'' particles.
In particular, the second KK photon $\gamma^{(2)}$ and Z boson $Z^{(2)}$ 
play an important role for the search of the second KK particles. They are 
able to decay into two charged leptons through the KK number violation 
operators. It is possible to reconstruct their masses clearly from the charged 
dileptons. Connecting their masses and those of the first KK particles, we can 
confirm the realization of the MUED model. Hence, $\gamma^{(2)}$ and $Z^{(2)}$ 
are the key ingredients for the discrimination. We therefore calculate their 
production rates at the LHC. This proceedings is based on the work \cite{Matsumoto:2009tb}.

\vspace{-3mm}

\section{Productions of $\gamma^{(2)}$ and $Z^{(2)}$} 

In this section, we discuss the Lagrangian relevant to the productions
of $\gamma^{(2)}$ and $Z^{(2)}$ bosons. With the Lagrangian we calculate 
the branching ratios of $g^{(2)}$ and $q^{(2)}$, which are necessary for the 
discussion of the indirect production of $\gamma^{(2)}$ and $Z^{(2)}$.

\subsection{Lagrangian for $\gamma^{(2)}$ and $Z^{(2)}$ productions} 

Firstly, we show the Lagrangian conserving the KK number relevant to gauge bosons, 

\begin{equation}
 \begin{split}
    \mathcal{L}_{\text{con}} 
    =
    &- g_i \sum_{n=1}^{\infty}
    \biggl[ 
          \bar f^{(n)} t^a \gamma^\mu f^{(n)} V_{i \mu}^{(0)a}    \\
    &~~~ + \bar f^{(n)} t^a \gamma^\mu P_{L(R)} f^{(0)} V_{i \mu}^{(n)a}
       + \bar f^{(0)} t^a \gamma^\mu P_{L(R)} f^{(n)} V_{i \mu}^{(n)a}   
    \biggr]   \\ 
    & -  \frac{g_i}{\sqrt{2}} \sum_{n, m=1}^{\infty}
    \biggl[ 
          \bar f^{(n)} t^a \gamma^\mu \gamma^5 f^{(m)} V_{i \mu}^{(n+m)a}    \\
    &~~~ + \bar f^{(n+m)} t^a \gamma^\mu f^{(n)} V_{i \mu}^{(m)a}
       + \bar f^{(n)} t^a \gamma^\mu f^{(n+m)} V_{i \mu}^{(m)a}    
    \biggr]   \\
    & + g_i f_i^{abc} \sum_{n=1}^{\infty}
    \biggl[
          (\partial_\mu V_{i \nu}^{(0)a}) V_i^{(n) b \mu} V_i^{(n) c \nu}   \\
       &~~~    
       + (\partial_\mu V_{i \nu}^{(n)a}) V_i^{(n) b \mu} V_i^{(0) c \nu}
       + (\partial_\mu V_{i \nu}^{(n)a}) V_i^{(0) b \mu} V_i^{(n) c \nu}   
    \biggr] ~,        
 \end{split}
\end{equation}
where the summation over $i$, $a$, $b$, $c$ are implicitly made. 
$f^{(n)}$, $g_i$,
$t^a$, $V_{i \mu}^{a}$ are listed in Table 1. In the third part of the
Lagrangian, $f_i^{abc}$ is the structure constant of $SU(3)$ for
gluon and that of
$SU(2)$ for W boson. As long as we use the KK
number conserving Lagrangian, second KK particles are produced in
pair due to the KK number conservation, and hence the production rates are
suppressed due to their small phase spaces.

\begin{table}
\tbl{$f^{(n)}$, $g_i$, $t^a$, $V_{i \mu}^{a}$ in the KK number conserving Lagrangian. $B$, $W$, $g$ 
                      are $U(1)$, $SU(2)$, $SU(3)$ gauge bosons, $g'$, $g_2$, $g_s$ are $U(1)$, $SU(2)$, 
                      $SU(3)$ gauge coupling constants, and $\sigma^a$, $\lambda^a$ are Pauli matrices, 
                      Gell-Mann matrices respectively.}
{\begin{tabular}{@{}lcccc@{}}
\toprule
  ~~~~~ n-th KK fermion $f^{(n)}$              & ~~~~~~$g_i$~~~~~~ & ~~~~~~$t^a$~~~~~~ & ~~~~~~$V_{i \mu}^{a}$~~~~~~ \\
\colrule
$SU(2)$-singlet charged lepton $E^{(n)}$   &   $- g'$              & $1$                    & $B_\mu$  \\
\hline
$SU(2)$-doublet lepton $L^{(n)}$              & $- (1/2) g'$       & $1$                    & $B_\mu$    \\
                                                         &        $g_2$        & $\sigma^a/2$       & $W^a_\mu$   \\
\hline
$SU(2)$-singlet up-type quark $U^{(n)}$    & $(2/3) g'$          & $1$                    & $B_\mu$ \\
                                                         &  $g_s$               & $\lambda^a/2$     & $g^a_\mu$      \\
\hline
$SU(2)$-singlet down-type quark $D^{(n)}$ & $- (1/3) g'$        & $1$                    & $B_\mu$ \\
                                                         &  $g_s$               & $\lambda^a/2$     &  $g^a_\mu$  \\
\hline
$SU(2)$-doublet quark $Q^{(n)}$              & $(1/6) g'$          & $1$                    & $B_\mu$ \\
                                                         &     $g_2$            &  $\sigma^a/2$     & $W^a_\mu$ \\
                                                         &     $g_s$            & $\lambda^a/2$     & $g^a_\mu$  \\
\botrule
\end{tabular}}
\label{table}
\end{table}

Next we discuss the KK number violating interactions. 
The effective Lagrangian for the KK number violating operators
turns out to be :
\begin{equation}
    \mathcal{L}_{\text{vio}}
    = 
    \frac{x_i}{4} 
    \Biggl\{
       ~N_i(f) c_t + 
       \left[ 
          9 C_j(f) - \frac{23}{3} C_j(G) \delta_{ij} + \frac{n_j}{3} \delta_{ij} 
       \right]
       c_j~
    \Biggr\}
    \bar f^{(0)} t_i^a \gamma^\mu P_{L (R)} f^{(0)} V_{i \mu}^{(2)a} ~,     \label{Lvio}
\end{equation}
\begin{equation}
 \begin{split}
    c_j \equiv \frac{ \sqrt{ 2 } x_j^2 }{ 16 \pi^2 } \text{log}\frac{ \Lambda^2 }{ \mu^2 } ~, ~~
    c_t \equiv \frac{ \sqrt{ 2 } y_t^2 }{ 16 \pi^2 } \text{log}\frac{ \Lambda^2 }{ \mu^2 }  ~.
 \end{split}       \label{c_i}
\end{equation}
Here 
$y_t$ is the top Yukawa coupling constant, $x_i$, $n_i$, $C_i(f)$, $C_i(G)$,
$t_i^a$ are listed in Table 2, and $N_i(f)$ is listed in Table 3. Indices
$i$, $j$ run over the SM gauge interactions $U(1)$, $SU(2)$, and
$SU(3)$, and summation over $f$ is implicitly made. The renormalization scale
is denoted by $\mu$.

\vspace{-3mm}

\begin{table}
\tbl{Coefficient in the KK number violating operator in the effective Lagrangian (Eq. (\ref{Lvio})). $g'$, $g_2$, 
                      $g_s$ are $U(1)$, $SU(2)$, $SU(3)$ gauge coupling constants, and $Y_f$ is 
                      $U(1)$ hypercharge. $\sigma^a$, $\lambda^a$ are Pauli and Gell-Mann matrices, respectively.}
{\begin{tabular}{@{}ccccc@{}}
\toprule
~~~~~~~~~~~~~~~~~~~~~~~~~~& ~~~~~~~~~~$U(1)$~~~~~~~~~~ & ~~~~~~~~~~$SU(2)$~~~~~~~~~~ & ~~~~~~~~~~$SU(3)$~~~~~~~~~~  \\
\colrule
$x_i$ & $g' Y_f$ & $g_2$ & $g_s$  \\
$n_i$ & 1 & 2 & 0 \\
$C_j(f)$ & $Y_f^2$ & 3/4 & 4/3 \\
$C_i(G)$ & 0 & 2 & 3 \\
$t_i^a$ & \textbf{1} & $\sigma^a/2$ & $\lambda^a/2$ \\
\botrule
\end{tabular}}
\label{tab:a}
\end{table}

\vspace{-5mm}

\begin{table}
\tbl{Coefficient $N(f)$ in the KK number violating operator in the effective Lagrangian (Eq. (\ref{Lvio})). $Q_3$ 
                      is the third generation $SU(2)$-doublet quark, and $T$ is $SU(2)$ singlet top quark.}
{\begin{tabular}{@{}ccccccc@{}}
\toprule
~~~& $Q_3^{(0)} Q_3^{(0)} \gamma^{(2)}$
     & $T^{(0)} T^{(0)} \gamma^{(2)}$
     & $Q_3^{(0)} Q_3^{(0)} W^{(2)}$
     & $Q_3^{(0)} Q_3^{(0)} g^{(2)}$
     & $T^{(0)} T^{(0)} g^{(2)}$         
     & other  \\
\colrule
$N(f)$     & ~~$1$~~ & ~~$5$~~ & ~~$1$~~ & ~~$1$~~ & ~~$1$~~ & 0   \\
\botrule
\end{tabular}}
\label{tab:b}
\end{table}

\vspace{-8mm}

\subsection{Production processes of $\gamma^{(2)}$ and $Z^{(2)}$} 

We are now in a position to discuss the production processes of
$\gamma^{(2)}$ and $Z^{(2)}$. At the LHC, $\gamma^{(2)}$ and $Z^{(2)}$
are produced through two type of processes : (1) direct productions and (2)
indirect productions via the cascade decays of the second KK colored
particles. The production cross sections of $\gamma^{(2)}$ and $Z^{(2)}$ has 
originally been calculated in Ref. 6, and their calculation 
includes all of the KK number conserving processes and the direct 1-body 
production processes of the second KK gauge bosons. In this article, we 
calculate the production cross sections of $\gamma^{(2)}$ and $Z^{(2)}$ 
including all significant processes. For example, our calculation includes 
$pp \rightarrow q^{(2)} q^{(0)}$, $pp \rightarrow \gamma^{(2)} q^{(0)}$, 
$pp \rightarrow q^{(2)} \bar q^{(0)}$ and so on. Importantly, these 
processes provide large contributions to $\gamma^{(2)}$ and $Z^{(2)}$ 
productions, particularly for large $1/R$ ($\gtrsim$ 800 GeV). We show the 
relevant processes to the $\gamma^{(2)}$ production through KK number 
violating interactions in Figure 1 - 3. We skip the discussion of the 
$\gamma^{(2)}$ production through KK number conserving interactions.
The processes of $Z^{(2)}$ production are almost same as the $\gamma^{(2)}$ 
production, so that we can skip the discussion of the $Z^{(2)}$ production.

\begin{figure}[t]
\begin{center}
\psfig{file=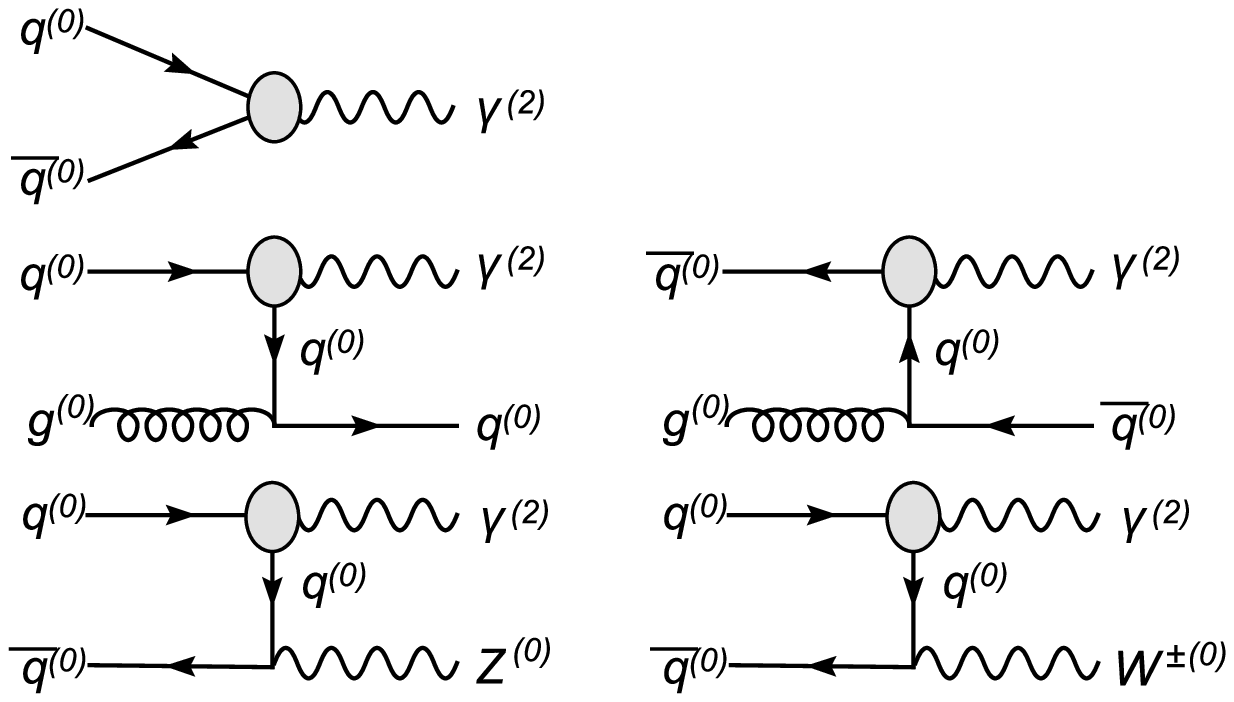,width=2.15in}
\end{center}
\caption{The direct production of  $\gamma^{(2)}$ through KK number violating processes.
                      The colored circle stands for the KK number violating vertex.}
\label{qq_to_gamma2}
\begin{center}
\psfig{file=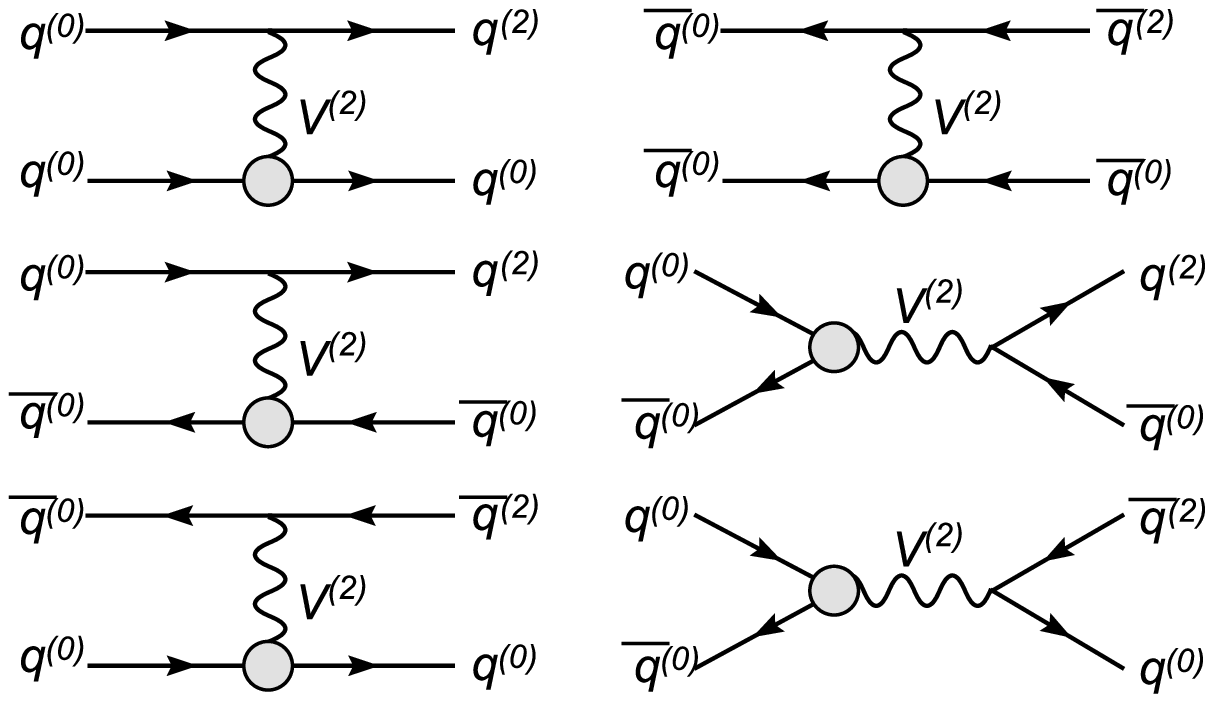,width=2.28in}
\end{center}
\caption{The production of $q^{(2)}$ through KK number violating processes.
            $V^{(2)}$ stands for $\gamma^{(2)}$, $W^{\pm (2)}$, $Z^{(2)}$, and $g^{(2)}$.
            The colored circle stands for the KK number violating vertex.}
\label{q2_pro_violating}
\begin{center}
\psfig{file=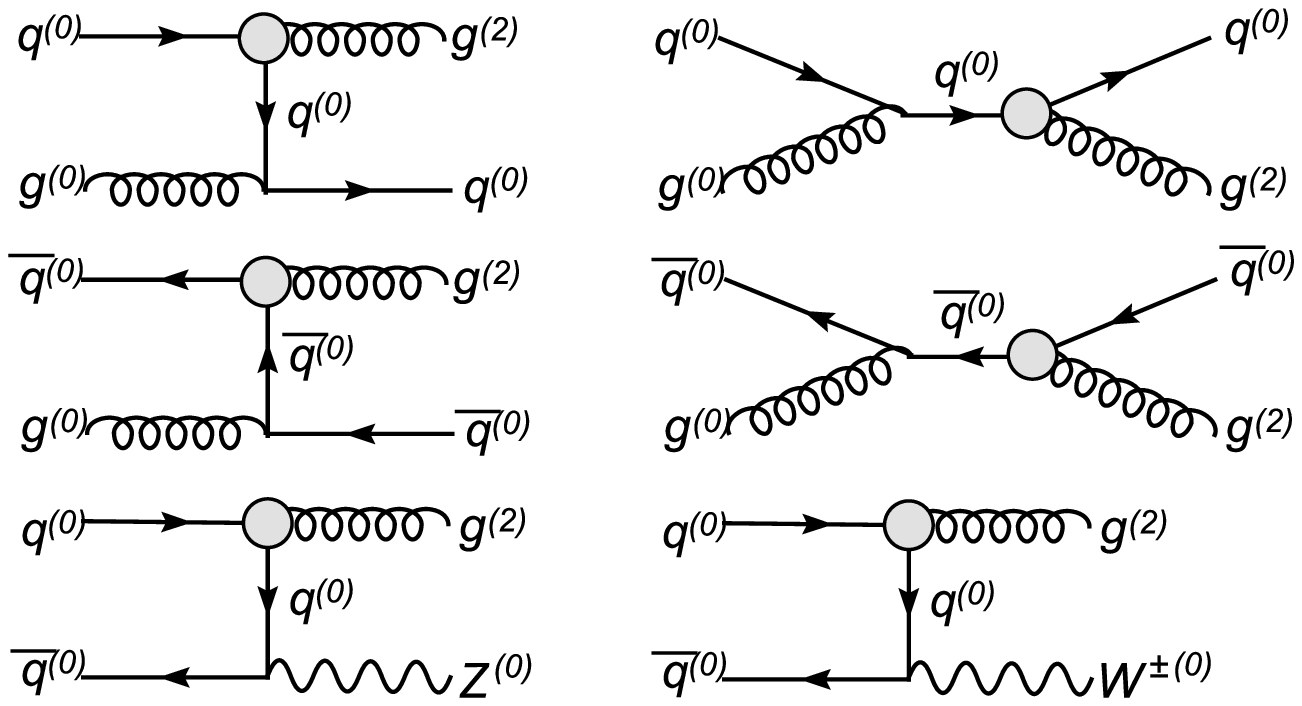,width=2.39in}
\end{center}
\caption{The production of $g^{(2)}$ through KK number violating processes.
            The colored circle stands for the KK number violating vertex.}
\label{g2_pro_violating}
\end{figure}

\section{Numerical results} 

\begin{figure}
\begin{center}
\psfig{file=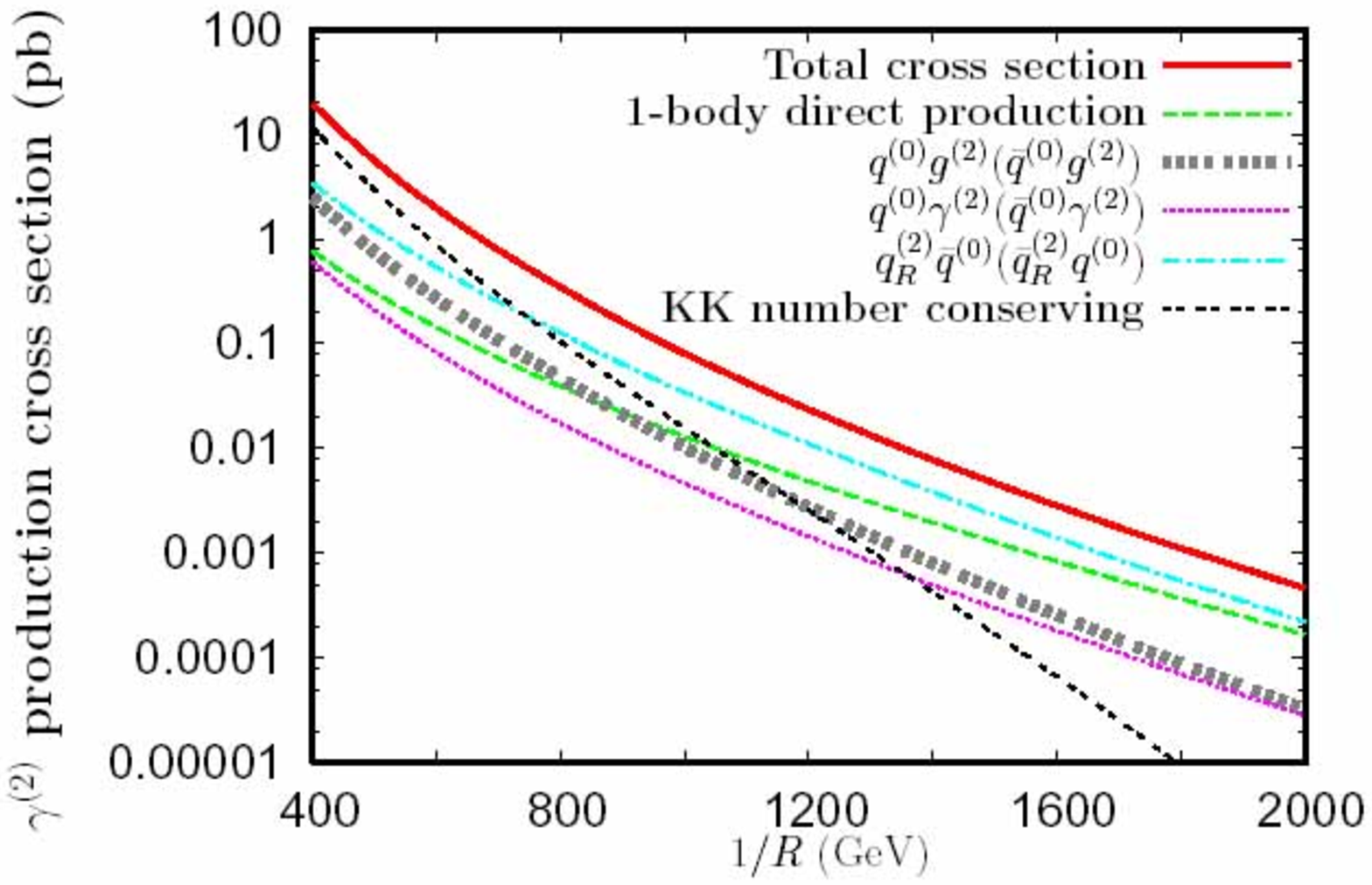,width=3.5in}
\end{center}
\caption{The production cross sections of $\gamma^{(2)}$. 
Red solid line shows the total cross section, and each line
shows partial cross section for each process as denoted in the legend.}
\label{gamma2_each_process}
\begin{center}
\psfig{file=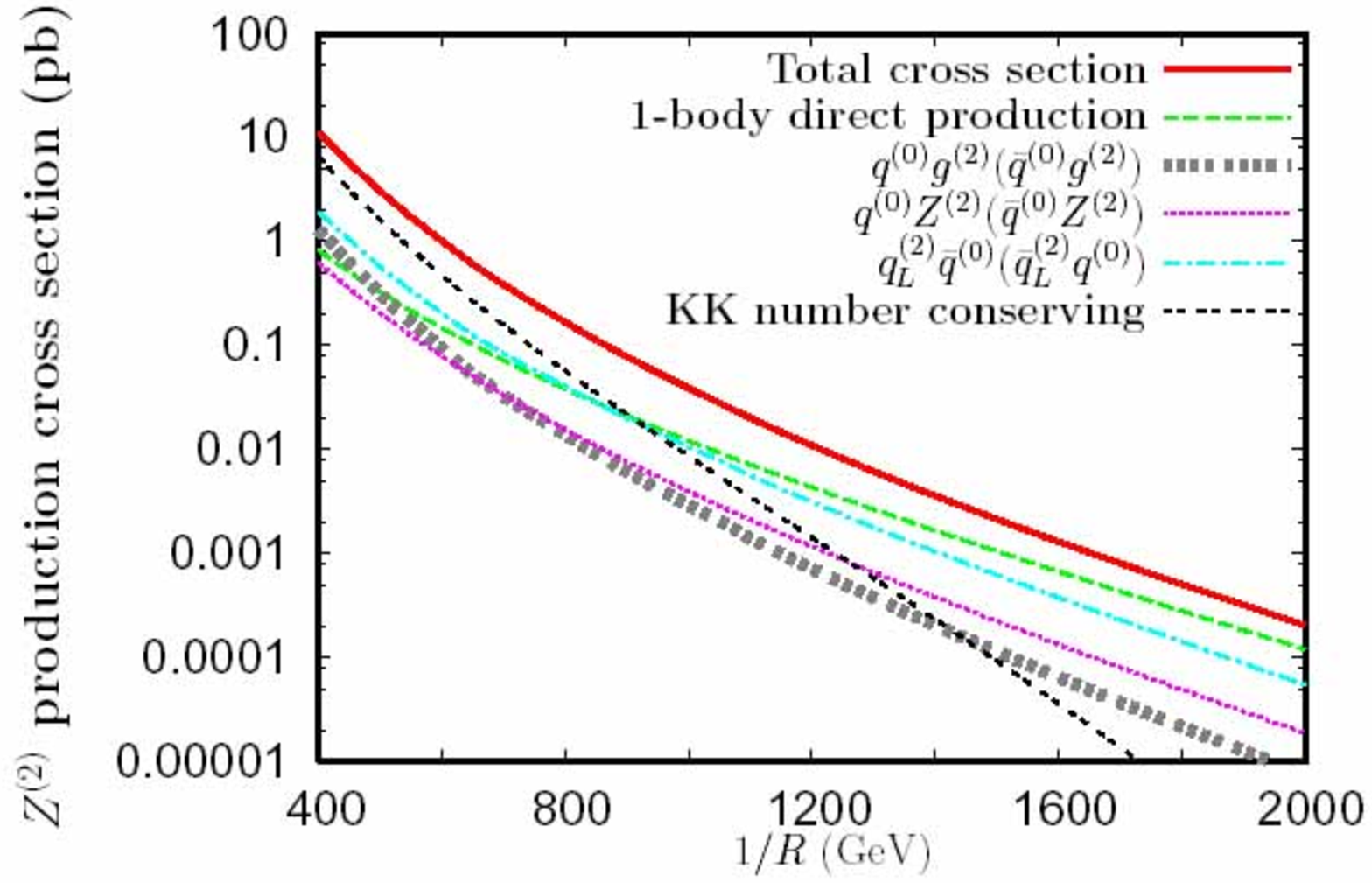,width=3.5in}
\end{center}
\caption{The production cross sections of $Z^{(2)}$. 
Red solid line shows the total cross section, and each line
shows partial cross section for each process as denoted in the legend.}
\label{Z2_each_process}
\end{figure}

In Figure \ref{gamma2_each_process} (Figure \ref{Z2_each_process}), we show the
production cross section of $\gamma^{(2)}$ ($Z^{(2)}$) as a function of $1/R$. 
In the calculation, we have used the CTEQ6L code \cite{Pumplin:2002vw} as a 
parton distribution function, the calculations of the cross sections have been 
performed by using the calcHEP \cite{Pukhov:2004ca} implementing the Lagrangian 
Eq. (\ref{Lvio}) derived in the previous section. Red solid line shows the total cross 
section of $\gamma^{(2)}$ ($Z^{(2)}$) production.

Assuming an integrated luminosity of 100 fb$^{-1}$, $\gamma^{(2)}$ and 
$Z^{(2)}$ will be produced 10$^6$ - 10$^2$ for 400 GeV $\leq 1/R \leq$ 
2000GeV. Once $\gamma^{(2)}$ and $Z^{(2)}$ are produced, they decay 
into dileptons with about 1\% branching ratios. 
In Table \ref{dilepton}, we show the number of the dilepton signals with assuming 
the luminosity 100 fb$^{-1}$. For $1/R \lesssim$ 1600 GeV, dilepton signals 
will be observed at least 1 event. If the MUED model is realized by the nature, 
in addition to the dilepton signals from the decay of
$\gamma^{(2)}$ and $Z^{(2)}$, many new particles with degenerate mass
spectrum around $1/R$, i.e., the first KK particles are
discovered. Connecting the observational results of dilepton signals and
the discovery of the first KK particles, it is possible to confirm the MUED model. 


\begin{table}
\tbl{The number of the dilepton signals at the LHC with 100 fb$^{-1}$.}
{\begin{tabular}{@{}cccc@{}}
\toprule
~~~~~~~~$1/R$~~~~~~~~ & ~~~~~dileptons from $\gamma^{(2)}$~~~~~ & ~~~~~dileptons from $Z^{(2)}$~~~~~ \\
\colrule
400 GeV & 1.5 $\times $ 10$^4$ & 9.4 $\times $ 10$^3$   \\
800 GeV & 2.9 $\times $ 10$^2$ & 1.6 $\times $ 10$^2$  \\
1200 GeV & 2.1 $\times $ 10 & 1.2 $\times $ 10  \\
1600 GeV & 2.6 & 1.4  \\
2000 GeV & 4.4 $\times $ 10$^{-1}$ & 2.3 $\times $ 10$^{-1}$  \\
\botrule
\end{tabular}}
\label{dilepton}
\end{table}


\section{Summary} 

At the LHC, the discrimination of the MUED model from other models is difficult, 
because the signals of new particles of TeV scale models are quite similar to 
each other. For the discrimination, we focused on the existence of the KK tower. 
Once $\gamma^{(2)}$ and $Z^{(2)}$ are produced, they can decay into dilepton 
which provides a very clean signal of the second KK particles. The discovery of 
the second KK particles, indicates the existence of the KK tower and hence will 
lead to the confirmation of the MUED model. In order to estimate the dilepton 
events from $\gamma^{(2)}$ and $Z^{(2)}$, we have calculated the production 
rates of $\gamma^{(2)}$ and $Z^{(2)}$ at the LHC.

We have calculated the KK number violating operators. They play a crucial role, 
because $\gamma^{(2)}$ and $Z^{(2)}$ can decay into dileptons through these 
couplings. Next we have shown all significant processes for $\gamma^{(2)}$ 
productions. Then we have calculated the production cross sections of 
$\gamma^{(2)}$ and $Z^{(2)}$. Finally we have found that the number of dilepton 
events are expected to be more than 1 for $1/R \lesssim$ 1600 GeV with an 
integrated luminosity of 100 fb$^{-1}$. We need further study to discuss the 
feasibility for the discovery of the MUED model by estimating the background 
from the SM. We leave it for a future work \cite{2009}.

\section*{Acknowledgments} 
 The work of J. S. was supported in part by the Grant-in-Aid for the Ministry of Education, Culture, Sports, Science, 
and Technology, Government of Japan (No. 20025001, 20039001, and 20540251). The work of M. Y. was supported 
in part by the Grant-in-Aid for the Ministry of Education, Culture, Sports, Science, and Technology, Government of 
Japan (No. 20007555).


\end{document}